\documentclass[10pt,twocolumn]{article}
\usepackage[top=2cm, bottom=2cm, left=1.8cm, right=1.8cm]{geometry}
\usepackage{times}  %
\usepackage[hyphens]{url}
\usepackage{graphicx}  %
\usepackage[keeplastbox]{flushend}
\usepackage{xcolor}
\usepackage{booktabs}
\usepackage{graphicx}
\usepackage{paralist}
\usepackage[small,bf]{caption}
\usepackage[tight]{subfigure}
\usepackage[hang,flushmargin]{footmisc}

\usepackage[compact]{titlesec}
\titlespacing*{\section}{0pt}{*4}{4pt}
\titlespacing{\subsection}{0pt}{*3}{3pt}
\usepackage{xspace}

\usepackage[hang,flushmargin]{footmisc}

\usepackage[compact]{titlesec}
\titlespacing*{\section}{0pt}{*4}{4pt} 
\titlespacing{\subsection}{0pt}{*3}{3pt}
\usepackage{xspace}

\usepackage{enumitem}
\setlist[itemize]{noitemsep, topsep=0pt}
\setlist[enumerate]{noitemsep, topsep=0pt}
\setlist[description]{noitemsep, topsep=0pt}

\usepackage{multirow}

\usepackage{amsmath,bm,amssymb}
\usepackage{calc}
\usepackage{wasysym}

\usepackage{tikz}
\usetikzlibrary{arrows}
\usepackage{pifont}

\newcommand{\fullcircle}{\ding{108}}
\newcommand{\emptycircle}{\ding{109}}
\newcommand{\halfcircle}{~\ding{119}}

\usepackage[bookmarks=true, bookmarksnumbered=true, colorlinks=true, linkcolor=linkcol, citecolor=citecol, urlcolor=urlcol, hypertexnames=true]{hyperref}

\definecolor{darkgreen}{RGB}{0, 100, 0}
\definecolor{linkcol}{rgb}{0.3,0,0}
\definecolor{citecol}{rgb}{0.3,0,0}
\definecolor{urlcol}{rgb}{0.3,0,0}

\makeatletter
\def\url@leostyle{%
  \@ifundefined{selectfont}{\def\UrlFont{\small}}%
  {\def\UrlFont{}}%
}
\makeatother
\usepackage{ifthen}
\newif\ifshort
\shorttrue   %
\ifshort
  \newcommand{\isShort}{true}
\else
  \newcommand{\isShort}{false}
\fi
\newcommand{\shortVer}[1]{\ifthenelse{\equal{\isShort}{true}}{{#1}}{}}
\newcommand{\longVer}[1]{\ifthenelse{\equal{\isShort}{false}}{{#1}}{}}

\newif\ifcomment
\commenttrue
\ifcomment
\newcommand{\sz}[1]{{\bf \textcolor{blue}{SZ: #1}}}
\newcommand{\jbnote}[1]{{\bf \textcolor{magenta}{JB: #1}}}
\newcommand{\jf}[1]{{\bf \textcolor{red}{JF: #1}}}

\else
\newcommand{\sz}[1]{}
\newcommand{\jbnote}[1]{}
\newcommand{\jf}[1]{}
\fi

\newcommand{\descr}[1]{\smallskip\noindent\textbf{#1}}

\begin{document}
\title{\bf{The Pushshift Reddit Dataset}}

\author{\bf Jason Baumgartner\textsuperscript{\rm 1,*}, Savvas Zannettou\textsuperscript{\rm 2,\smiley{}}, Brian Keegan\textsuperscript{\rm 3}, Megan Squire\textsuperscript{\rm 4}, Jeremy Blackburn\textsuperscript{\rm 5,\smiley{}}\\[0.5ex] %
\normalsize \textsuperscript{\rm 1}Pushshift.io, \textsuperscript{\rm 2}Max Plank Institute, \textsuperscript{\rm 3} University of Colorado Boulder, \textsuperscript{\rm 4}Elon University, \textsuperscript{\rm 5}Binghamton University\\
\normalsize \textsuperscript{*}Network Contagion Research Institute, \textsuperscript{\smiley{}}iDRAMA Lab\\
\normalsize jason@pushshift.io, szannett@mpi-inf.mpg.de, brian.keegan@colorado.edu, msquire@elon.edu, blackburn@cs.binghamton.edu
}
\date{}

\maketitle

\begin{abstract}
Social media data has become crucial to the advancement of scientific understanding.
However, even though it has become ubiquitous, just collecting large-scale social media data involves a high degree of engineering skill set and computational resources.
In fact, research is often times gated by data engineering problems that must be overcome before analysis can proceed.
This has resulted recognition of datasets as meaningful research contributions in and of themselves.

Reddit, the so called ``front page of the Internet,'' in particular has been the subject of numerous scientific studies.
Although Reddit is relatively open to data acquisition compared to social media platforms like Facebook and Twitter, the technical barriers to acquisition still remain.
Thus, Reddit's millions of subreddits, hundreds of millions of users, and hundreds of billions of comments are at the same time relatively accessible, but time consuming to collect and analyze systematically.

In this paper, we present the Pushshift Reddit dataset.
Pushshift is a social media data collection, analysis, and archiving platform that since 2015 has collected Reddit data and made it available to researchers.
Pushshift's Reddit dataset is updated in real-time, and includes historical data back to Reddit's inception.
In addition to monthly dumps, Pushshift provides computational tools to aid in searching, aggregating, and performing exploratory analysis on the entirety of the dataset.
The Pushshift Reddit dataset makes it possible for social media researchers to reduce time spent in the data collection, cleaning, and storage phases of their projects.

\end{abstract}

\section{Introduction}\label{sec:intro}

Understanding complex socio-technical phenomena requires data-driven research based on large-scale, reliable, relevant data sets.
Web data, particularly data from application programming interfaces (APIs), has been an enormous boon for researchers using online social platforms' databases of user-generated activity and content~\cite{Freelon_InterpretationDigitalTrace_2014,GolderDigitalFootprintsOpportunities2014,Hampton_StudyingDigitalDirections_2017,Lazer_DataexMachina_2017}.
The ability to ``crawl'' and ``scrape'' large-scale and high-resolution samples of publicly-accessible user data stimulated emerging fields like social computing~\cite{wang_social_2007} and computational social science~\cite{LazerComputationalSocialScience2009}, and developed new fields like crisis informatics~\cite{Palen_CrisisinformaticsNew_2016}.
But following major scandals around data privacy and ethics, social media platforms like Facebook and Twitter changed previously permissive data access provisions of their public APIs~\cite{Walker_disinformationlandscapelockdown_2019}.
As a consequence, the ability for researchers to collect timely data, share tools, instruct students, and reproduce findings has been curtailed.

This ``post-API age'' is characterized by the deprecation of data resources used for research and teaching~\cite{Freelon_ComputationalResearchPostAPI_2018,Puschmann_endwildwest_2019}, increased stratification of data access based on social, technical, and financial capital~\cite{boydCriticalQuestionsBig2012,manovich2011trending}, and greater fear of prosecution around violating terms of service in the course of research~\cite{Halavais_Overcomingtermsservice_2019,Patel_TestingLimitsFirst_2018}.
These changes have had a profoundly chilling effect on researchers' use of API-derived data to investigate behavior like discrimination, harassment, radicalization, hate speech, and disinformation. Furthermore, researchers have struggled in systematically studying the role that platforms' changing features, design affordances, and governance strategies play in sustaining these forms of ``turpitude-as-a-service''~\cite{Bruns_APIcalypsesocialmedia_2019,Keegan_DiscoveringSocial_2018}.
Faced with conflicting incentives between protecting their users' data from abuse and maintaining their commitments to values of openness, online social platforms are exploring alternative data sharing models like ``trusted third party'' models that still carry significant technical and reputational risks~\cite{Bruns_APIcalypsesocialmedia_2019,gibney_privacy_2019,ingram_silicon_2019,Mervis_Privacyconcernscould_2019,Puschmann_endwildwest_2019}.

Even if the ``golden age'' of API-driven computational social science and social computing research had not closed in the shadow of privacy scandals, it was nevertheless characterized by enormous inefficiencies in data collection and inequalities in access~\cite{manovich2011trending,Puschmann_endwildwest_2019}, ethically-suspect methods and implications~\cite{BoydUntanglingresearchpractice2016,tufekci2014big,Olteanu_SocialDataBiases_2019}, a lack of concern for data sharing or reproducibility~\cite{Borgman_conundrumsharingresearch_2012,Weller_manifestodatasharing_2016}, and failures to validate constructs or generalize to off-platform behavior~\cite{Ekbia_Bigdatabigger_2015,howison_validity_2011,Japec_BigDataSurvey_2015}.
Facebook's and Twitter's changes in data access were significant, however the enclosure of previously open big social data sources is not ubiquitous among platform providers~\cite{boyle2017second,hess2003ideas,hunter2003cyberspace}.
Social platforms and online communities like Wikipedia~\cite{wikimedia_dump_2019}, Stack Exchange~\cite{stackexchange_dump_2019}, GitHub~\cite{Gousi13}, and Reddit~\cite{reddit_api_2019} continue to offer open APIs and data dumps that are valuable for researchers.

In this paper, we assist to the goal of providing open APIs and data dumps to researchers by releasing the Pushshift Reddit dataset.
In addition to monthly dumps of 651M submissions and 5.6B comments posted on Reddit between 2005 and 2019\footnote{Available at \url{https://files.pushshift.io/reddit/}}, the Pushshift Reddit dataset also includes an API for researcher access and a Slackbot that allows researchers to easily interact with the collected data. 
The Pushshift Reddit API enables researchers to easily execute queries on the whole dataset without the need for downloading the monthly dumps.
This reduces the requirement for substantial storage capacity, thus making the data more available to a wider range of users.
Finally, we provide access to a Slackbot that allows researchers to easily produce visualizations of data from the Pushshift Reddit dataset in real-time and discuss them with colleagues on Slack.
These resources allow research teams to quickly begin interacting with data with very little time spent on the tedious aspects of data collection, cleaning, and storage.

\section{Pushshift}

Pushshift is not a new or isolated data platform, but a five year-old platform with a track record in peer-reviewed publications and an active community of several hundred users.
Pushshift not only collects Reddit data, but exposes it to researchers via an API.
Why do people use Pushshift's API instead of the official Reddit API? 
In short, Pushshift makes it much easier for researchers to query and retrieve historical Reddit data, provides extended functionality by providing full-text search against comments and submissions, and has larger single query limits. 
Specifically, because, at the time of this writing, Pushshift has a size limit five times greater than Reddit's 100 object limit, Pushshift enables the end user to quickly ingest large amounts of data. 
Additionally, the Pushshift API offers aggregation endpoints to provide summary analysis of Reddit activity, a feature that the Reddit API lacks entirely.

The Pushshift Reddit dataset provides not just a \textit{technical} infrastructure of software and hardware for collecting ``big social data'' but also a \textit{social} infrastructure of organizational processes for responsibly collecting, governing, and discussing these research data.

\subsection{Data collection process}
\begin{figure}[t]
    \centering
    \includegraphics[width=\columnwidth]{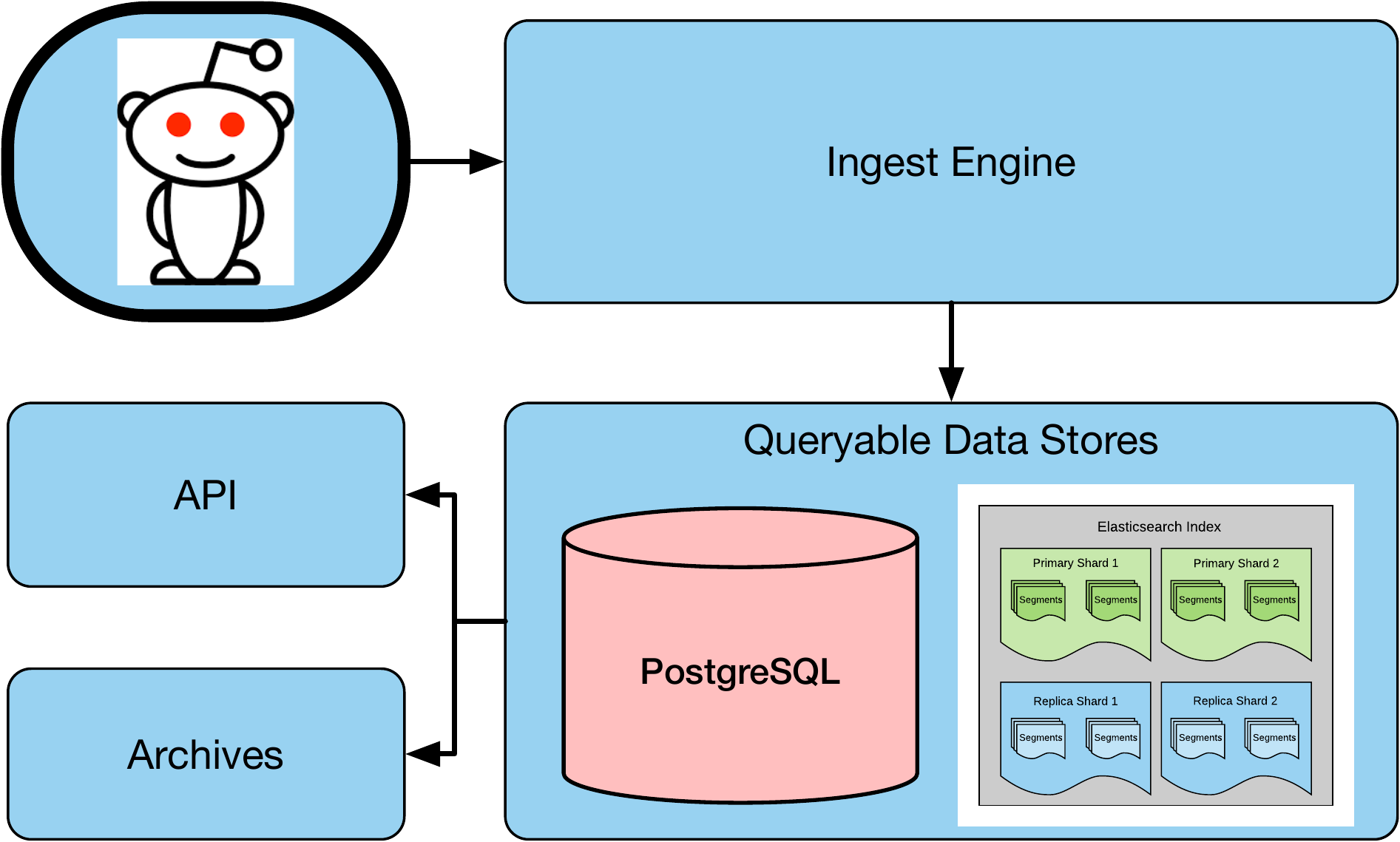}
	\caption{Pushshift's Reddit data collection platform.}
	\label{fig:architecture}
\end{figure}

Pushshift uses multiple backend software components to collect, store, catalog, index, and disseminate data to end-users.
As seen in Fig.~\ref{fig:architecture}, these subsystems are:
\begin{itemize}
	\item The \textbf{ingest engine}, which is responsible for collecting and storing raw data.
	\item A PostgreSQL \textbf{database}, which allows for advanced querying of data and meta-data storage.
	\item An Elastic Search \textbf{document store} cluster, which performs indexing and aggregation of ingested data.
	\item An \textbf{API} to allow researchers dynamic access to collected data and aggregation functionality.
\end{itemize}

\paragraph{Ingest Engine.}
The first stage in the Pushshift pipeline is the ingest engine, which is responsible for actually collecting data.
The ingest engine can be thought of as a \emph{framework} for large scale collection of heterogeneous social media data sources.
The ingest engine orchestrates the execution of a multiple data collection programs, each designed to handle a particular data source.
Specifically, the ingest engine provides and manages a job scheduling queue, and provides a set of common APIs to handle the data storage. Currently, Pushshift's ingest engine works as follows:

First, the program runner starts each ingest program (\textit{i.e.}, the programs that actually collect the data).
The ingest engine is agnostic to the particulars of the individual ingest programs: no particular programming language is required, and there is no particular expectation of how an ingest program works, modulo its interactions with the remainder of the ingest engine.
Typically, an ingest program will directly interact with Web APIs, scrape content from HTML pages, use data streams where available, \textit{etc}.
Next, the ingest program inserts the raw data retrieved into a database as well as into a document store.
Behind the scenes, each piece of collected data is added to an intermediate queue (currently implemented via Redis), which serves as a staging area until the data is processed by any custom processing scripts the ingest program's creator might require.
Finally, the raw data is periodically flushed to disk.
The data storage format can be specified by the ingest program creator via the custom processing scripts previously mentioned, or a standard, Pushshift-implemented format can be used (\textit{e.g.}, \texttt{ndjson}).

\paragraph{PostgreSQL \& ElasticSearch.}
Pushshift currently uses Elasticsearch (ES) as a scalable document store for each data source that is part of the ingest pipeline.
ES offers a number of important features for storing and analyzing large amounts of data.
For example, ES achieves \emph{ease-of-scaling} by utilizing a cluster approach for horizontal expansion. It ensures \emph{redundancy} by creating multiple replicas for each index so that a node outage does not affect the overall health of the cluster. The ES robust dynamic mapping tools allow \emph{easy modification and expansion} of indexes to accommodate changes in data structure from the source.
This is useful because Reddit’s API does not implement any type of versioning, yet there are constant additions and modifications made to the API when new features and data types are added to the response objects.
By using dynamic mapping types, Pushshift can easily add new fields to existing indices.
This enables us to quickly modify the corresponding mappings to allow search and aggregation on those new fields.
Pushshift also makes use of the ICU Analysis plug-in for ES~\cite{ICU_2019,ICU_ES_2019}, which provides support for international locales, full Unicode support up through Unicode 12, and complete emoji search support.

\descr{API} Pushshift currently allows users to search Reddit data via an API.
Right now, this API exports much of the search and aggregation functionality provided by Elastic Search.
This functionality supports dozens of community applications and numerous research projects.
The API is the major workload of handled by Pushshift's computational resources, serving 500M requests per month.
Although in this paper we focus on a description of the data (Section~\ref{sec:data}) due to space limitations, we provide online API documentation at \url{https://pushshift.io/api-parameters/}.

\begin{figure}[t]
    \centering
    \includegraphics[width=\columnwidth]{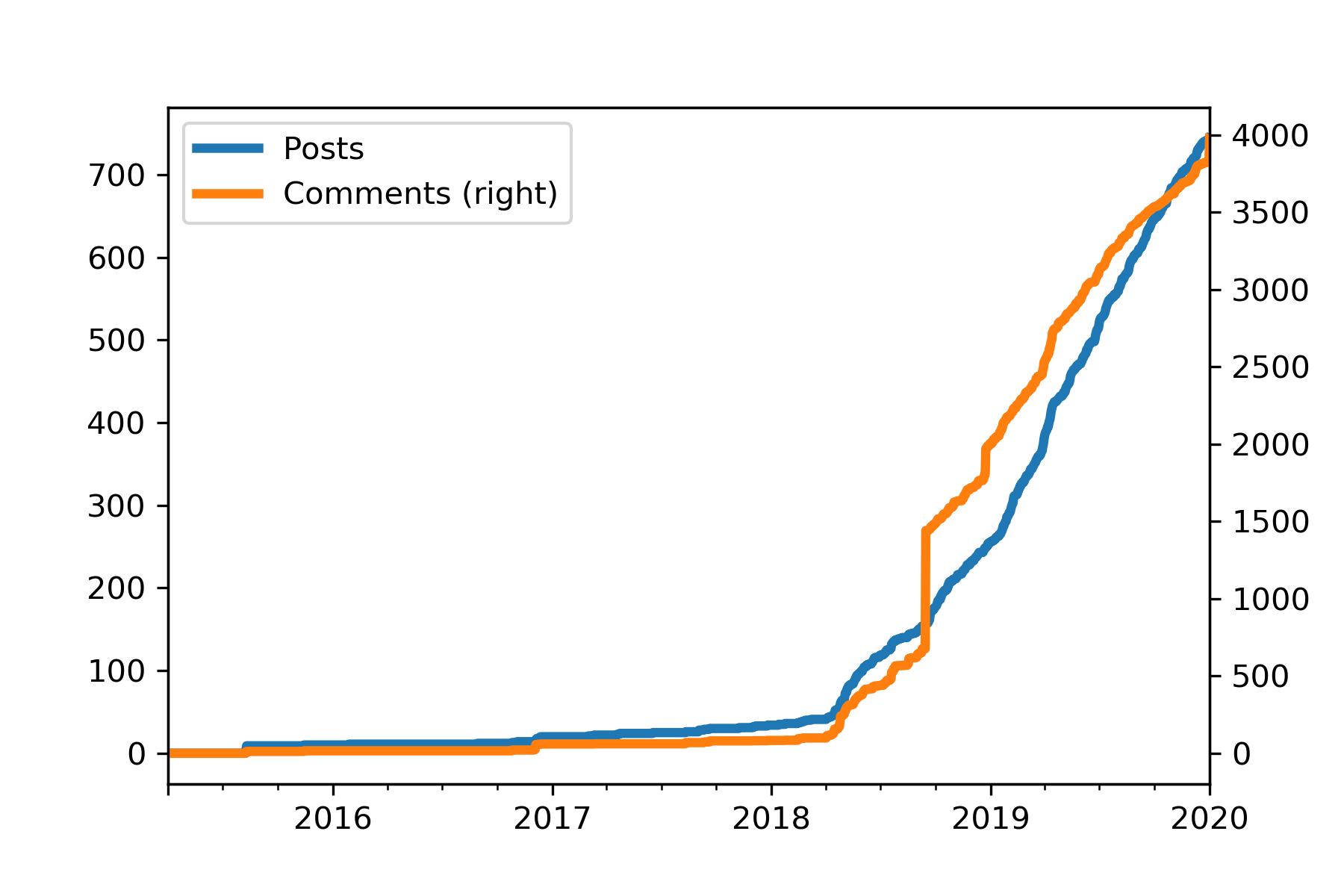}
    \caption{Activity on the r/pushshift subreddit.}
    \label{fig:r-pushshift}
\end{figure}

\descr{Community}
In addition to Pushshift's website, which features an interactive dashboard of current activity trends, Pushshift also has two active user communities on Reddit and Slack. 
The /r/pushshift subreddit was created in April 2015 and is used for sharing announcements, answering questions, reporting bugs, and soliciting feedback for new features.
There are more than 2,100 subscribers to this subreddit, an active team of 10 moderators, and more than 700 posts (with more than 4,000 comments) from over 350 unique users (see Fig.~\ref{fig:r-pushshift}).

\begin{figure}[t]
    \centering
    \includegraphics[width=\columnwidth]{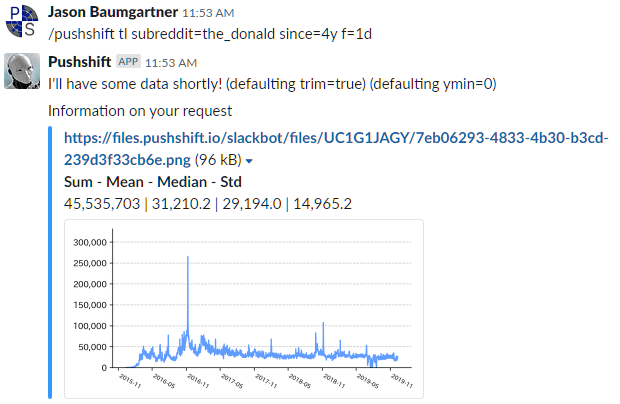}
	\caption{The Pushshift chatbot in Slack.}
	\label{fig:slack_chatbot}
\end{figure}

\begin{figure}[t!]
    \centering
    \includegraphics[width=\columnwidth]{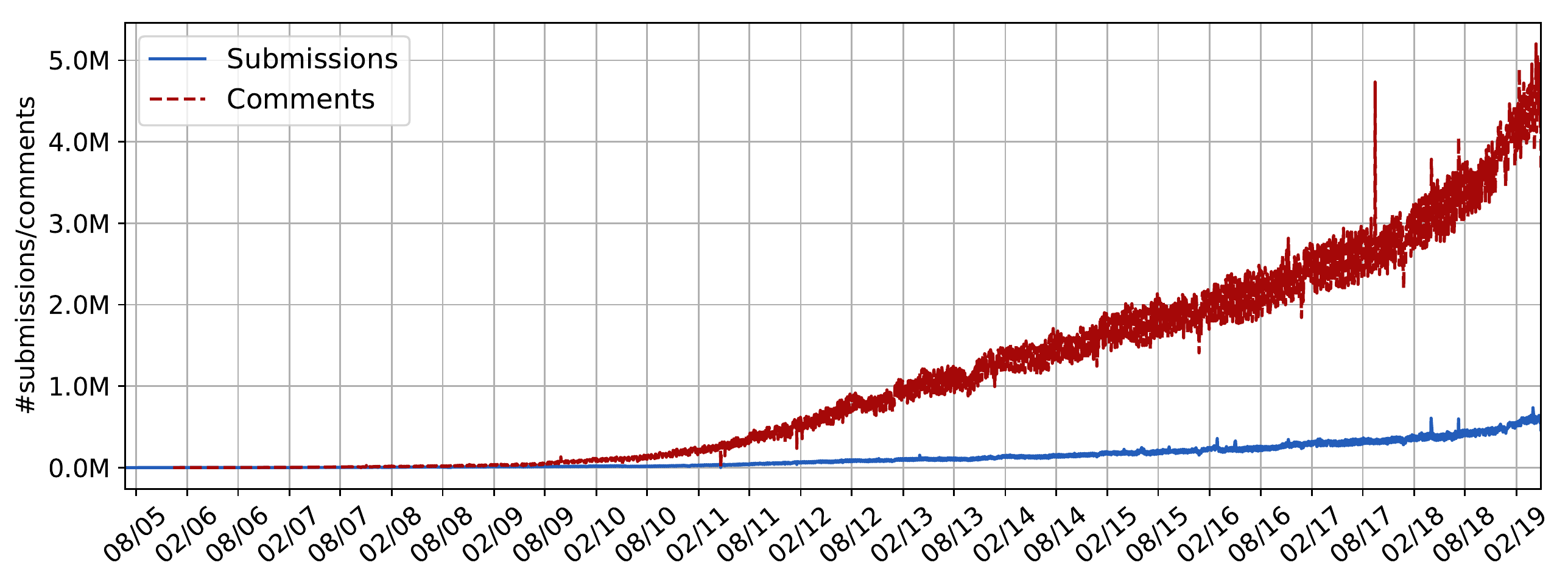}
	\caption{The number of submissions and comments for each day of our dataset.}
	\label{fig:counts_per_day}
\end{figure}

The Pushshift Slack team has nearly 300 registered users and more than 260,000 messages across 53 channels discussing data science and visualization.
Custom tools have also been developed to integrate the Pushshift archive into these Slack communities.
For example, users can interact with a Slack chatbot in realtime.
The bot can analyze and visualize Pushshift data based on queries made in the Slack channel, and return those visualizations to the channel for discussion and observation.
In Fig.~\ref{fig:slack_chatbot}, a user queried the total number of daily comments to the /r/the\_donald subreddit by day over the past four years and received a time series plot and summary statistics from the chatbot within a few seconds. 
The chatbot can also be shared to other non-Pushshift workspaces, allowing researchers in other Slack workspaces to use the data. 
This extends the reach of Pushshift data even further.

\begin{table*}[ht]
\centering

\begin{tabular}{|p{0.17\textwidth}|p{0.8\textwidth}|}
  \hline
 Field & Description \\ 
  \hline
   \hline
\textbf{id} & {The submission's identifier, e.g., ``5lcgjh'' (String).} \\ 
  \hline
\textbf{url} & The URL that the submission is posting. This is the same with the permalink in cases where the submission is a self post. E.g., ``\url{https://www.reddit.com/r/AskReddit/} \\ 
  \hline
\textbf{permalink} & Relative URL of the permanent link that points to this specific submission, e.g., ``{/r/AskReddit/comments/5lcgj9/what\_did\_you\_think\_of\_the\_ending\_of\_rogue\_one/}'' (String). \\ 
  \hline
\textbf{author} & The account name of the poster, e.g., ``example\_username'' (String). \\ 
  \hline
\textbf{created\_utc} & UNIX timestamp referring to the time of the submission's creation, e.g., 1483228803 (Integer). \\ 
  \hline
\textbf{subreddit} & Name of the subreddit that the submission is posted. Note that it excludes the prefix /r/. E.g., 'AskReddit' (String). \\ 
  \hline
\textbf{subreddit\_id} & The identifier of the subreddit, e.g., ``t5\_2qh1i'' (String). \\ 
  \hline
\textbf{selftext} & The text that is associated with the submission (String). \\ 
  \hline
\textbf{title} & The title that is associated with the submission, e.g., ``What did you think of the ending of Rogue One?'' (String). \\ 
  \hline
\textbf{num\_comments} & The number of comments associated with this submission, e.g., 7 (Integer). \\ 
  \hline
\textbf{score} & The score that the submission has accumulated. 
The score is the number of upvotes minus the number of downvotes. 
E.g., 5 (Integer).
\textbf{NB:} Reddit fuzzes the real score to prevent spam bots.  \\ 
\hline
\textbf{is\_self} & Flag that indicates whether the submission is a self post, e.g., true (Boolean). \\ 
  \hline
\textbf{over\_18} & Flag that indicates whether the submission is Not-Safe-For-Work, e.g., false (Boolean). \\ 
  \hline
\textbf{distinguished} & Flag to determine whether the submission is distinguished\footnote{See \url{https://www.reddit.com/r/redditdev/comments/19ak1b/api_change_distinguished_is_now_available_in_the/}} by 
moderators. ``null'' means not distinguished (String). \\ 
  \hline
\textbf{edited} & Indicates whether the submission has been edited. Either a number indicating the UNIX timestamp that the submission was edited at, ``false'' otherwise. \\ 
  \hline
\textbf{domain} & The domain of the submission, e.g., self.AskReddit (String). \\ 
  \hline
\textbf{stickied} & Flag indicating whether the submission is set as sticky in the subreddit, e.g., false (Boolean). \\ 
  \hline
\textbf{locked} & Flag indicating whether the submission is currently closed to new comments, e.g., false (Boolean). \\ 
  \hline
\textbf{quarantine} & Flag indicating whether the community is quarantine, e.g., false (Boolean). \\ 
  \hline
\textbf{hidden\_score} & Flag indicating if the submission's score is hidden, e.g., false (Boolean). \\ 
  \hline
\textbf{retrieved\_on} & UNIX timestamp referring to the time we crawled the submission, e.g., 1483228803 (Integer). \\ 
  \hline
\textbf{author\_flair\_css\_class} & The CSS class of the author's flair. This field is specific to subreddit (String). \\ 
  \hline
\textbf{author\_flair\_text} & The text of the author's flair. This field is specific to subreddit (String). \\ 
   \hline
\end{tabular}
\caption{\texttt{Submission}s data description.}
\label{tbl:data_description_submissions}
\end{table*} 

\section{Description of the Pushshift Reddit Dataset}\label{sec:data}

Pushshift makes available all the submissions and comments posted on Reddit between June 2005 and April 2019. 
The dataset consists of 651,778,198 submissions and 5,601,331,385 comments posted on 2,888,885 subreddits.
Fig.~\ref{fig:counts_per_day} shows the number of submissions and comments per day. 
We observe that the number of submissions and comments increase over the course of our dataset.
After August 2013, we have consistently over 1M comments per day, while by the end of our dataset (April 2019) we have 5M comments per day.
Also, while submissions are substantially fewer than comments, submissions have reached reached a consistent level of over 500K per day in this dataset.

The Pushshift Reddit dataset is made up of two sets of files: one set of files for the submissions and one for the comments.
Below, we describe the structure of each of the files in these two sets.

\descr{Submissions.} The submissions dataset consists of a set of newline delimited JSON\footnote{\url{http://ndjson.org/}} files: we maintain a separate file for each month of our data collection.
Each line in these files correspond to a submission and it is a JSON object.
Table~\ref{tbl:data_description_submissions} describes the most important key/values included in each submission's JSON object.

\descr{Comments.} Similarly to the submissions, the comments' dataset is a collection of ndjson files with each file corresponding to a month-worth of data.
Each line in these files correspond to a comment and it is a JSON object.
Table~\ref{tbl:data_description_posts} describes the most important keys/values in each comment's JSON object.

\begin{table*}[ht]
\centering

\begin{tabular}{|p{0.17\textwidth}|p{0.75  \textwidth}|}
  \hline
 Field & Description \\ 
 \hline
 \hline
 \textbf{id} & The comment's identifier, e.g., ``dbumnq8'' (String). \\
 \hline
\textbf{author} & The account name of the poster, e.g., ``example\_username'' (String).  \\
\hline
\textbf{link\_id} & Identifier of the submission that this comment is in, e.g., ``t3\_5l954r'' (String).  \\
\hline
\textbf{parent\_id} & Identifier of the parent of this comment, might be the identifier of the submission if it is top-level comment or the identifier of another comment, e.g., ``t1\_dbu5bpp'' (String).  \\
\hline
\textbf{created\_utc} & UNIX timestamp that refers to the time of the submission's creation, e.g., 1483228803 (Integer).  \\
\hline
\textbf{subreddit} & Name of the subreddit that the comment is posted. Note that it excludes the prefix /r/. E.g., 'AskReddit' (String).  \\
\hline
\textbf{subreddit\_id} & The identifier of the subreddit where the comment is posted, e.g., ``t5\_2qh1i'' (String).  \\
\hline
\textbf{body} & The comment's text, e.g., ``This is an example comment'' (String).  \\
\hline
\textbf{score} & The score of the comment.
The score is the number of upvotes minus the number of downvotes. 
Note that Reddit fuzzes the real score to prevent spam bots. E.g., 5 (Integer). \\
\hline
\textbf{distinguished} & Flag to determine whether the comment is distinguished by the moderators. ``null'' means not distinguished\footnote{See \url{https://www.reddit.com/r/redditdev/comments/19ak1b/api_change_distinguished_is_now_available_in_the/} for more details} (String).  \\
\hline
\textbf{edited} & Flag indicating if the comment has been edited. Either the UNIX timestamp that the comment was edited at, or ``false''.  \\
\hline
\textbf{stickied} & Flag indicating whether the submission is set as sticky in the subreddit, e.g., false (Boolean).  \\
\hline
\textbf{retrieved\_on} & UNIX timestamp that refers to the time that we crawled the comment, e.g., 1483228803 (Integer).  \\
\hline
\textbf{gilded} & The number of times this comment received Reddit gold, e.g., 0 (Integer).  \\
\hline
\textbf{controversiality} & Number that indicates whether the comment is controversial, e.g., 0 (Integer).  \\
\hline
\textbf{author\_flair\_css\_class} & The CSS class of the author's flair. This field is specific to subreddit (String).  \\
\hline
\textbf{author\_flair\_text} & The text of the author's flair. This field is specific to subreddit (String).  \\ 
\hline
\end{tabular}
\caption{\texttt{Comment}s data description.}
\label{tbl:data_description_posts}
\end{table*}

\descr{FAIR principles.} The Pushshift Reddit dataset aligns with the FAIR principles.\footnote{\url{https://www.go-fair.org/fair-principles/}}
Our dataset is \emph{Findable} as the monthly dumps are publicly available via Pushshift's website\footnote{\url{https://files.pushshift.io/reddit/}}.
We also upload a small sample of the dataset to the Zenodo service, so that we obtain a persistent digital object identifier (DOI): \href{https://zenodo.org/record/3608135}{10.5281/zenodo.3608135}.
Note that we were unable to upload the entire dataset to Zenodo, since the service has a limit of 100GB and our dataset is in the order of several terabytes.
The Pushshift Reddit dataset is \emph{Accessible} as it can be accessed by anyone visiting the Pushshift's website.
Furthermore, we offer an API and a Slackbot that allow researchers to easily execute queries and obtain data from our infrastructure without the need to download the large monthly dumps.
Also, our dataset is \emph{Interoperable} because it is JSON format, which is a widely known and used format for data.
Because the provenance for the collected data is very clear, and users are simply asked to cite Pushshift in order to use the data, our dataset is also \emph{Reusable}.

\section{Dataset Use Cases}

\begin{figure}[t!]
    \centering
    \includegraphics[width=0.80\columnwidth]{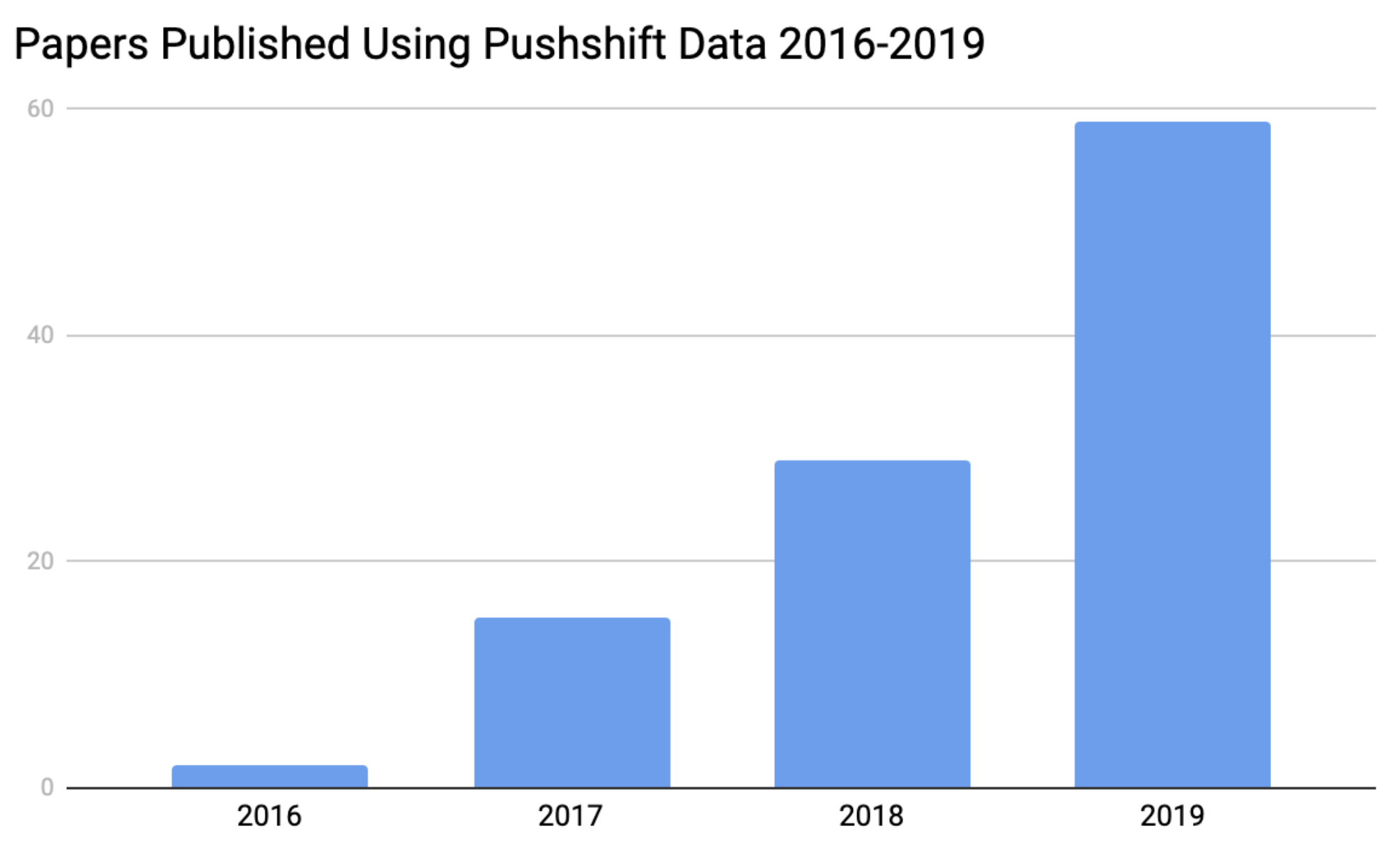}
    \caption{Over 100 peer-reviewed papers have been published using Pushshift data.}
    \label{fig:pushshift_paper_counts}
\end{figure}

The Pushshift Reddit dataset has attracted a substantial research community.%
As of late 2019, Google Scholar indexes over 100 peer-reviewed publications that used Pushshift data (see Fig.~\ref{fig:pushshift_paper_counts}).
This research covers a diverse cross-section of research topics including measuring toxicity, personality, virality, and governance.
Pushshift's influence as a primary source of Reddit data among researchers has attracted empirical scrutiny~\cite{gaffney_caveat_2018}, which in turn has led to improved data validation efforts~\cite{baumgartner_incompleteness_2018}.
We note that there is some difficulty in ascertaining our dataset's full contribution to the scientific community due to a previous lack of deliberate efforts to conform to FAIR principles, which we address in this paper.

\paragraph{Online community governance.} 
Reddit's ecosystem of sub-reddits are primarily governed by volunteer moderators with substantial discretion over creating and enforcing rules about user behavior and content moderation~\cite{fiesler2018reddit,squirrell_platform_2019}.
This distributed and volunteer-led model stands in contrast to the centralized strategies of other prominent social platforms like Facebook, Twitter, and YouTube~\cite{seering_moderator_2019}.
These differences between centralized versus delegated moderation make ideal case studies for comparing the effectiveness of responses to difficult issues like social movements, fringe identities, hate speech, and harassment campaigns~\cite{massanari_gamergate_2017,matias_dark_2016,matias_civic_2019}.
Pushshift data has already been instrumental for researchers exploring the spillover effects of banning offensive sub-communities~\cite{chandrasekharan_ban_2017}, identifying common features of abusive behavior across communities~\cite{chandrasekharan_bag_2017}, similarity in norms and rules across communities~\cite{chandrasekharan_hidden_2018,fiesler2018reddit}, perceptions of fairness in moderation decisions~\cite{jhaver_reactions_2019,jhaver_transparency_2019}, and improving automated moderation tools~\cite{chandrasekharan_crossmod_2019}.

\paragraph{Online extremism.} 
The political extremism research community currently faces significant challenges in understanding how mainstream and fringe online spaces are used by bad actors. 
Despite widespread agreement that recent increases in online radicalization are due to ``a globalised, toxic, anonymous online culture'' operating largely outside mainstream social media platforms~\cite{onlinehateprevention}, much of the research on extremist use of social media still focuses on mainstream sites like Facebook or Twitter~\cite{burris_and_strahm}. 
Access to these rapidly-changing online spaces is difficult, and many research teams end up using out-of-date data, or relying on the data they have, rather than the data they need. 
Many social media platforms face pressure to monetize their data~\cite{botta} or remove access to it entirely~\cite{bastos}, making research access to these spaces expensive and difficult. 
Yet, extremism researchers agree that data access is a key limitation to understanding online radicalization as a phenomenon. 
Online extremism researchers top recommendation is to ``invest more in data-driven analysis of far-right violent extremist and terrorist use of the Internet.''~\cite{tech_versus_terrorism} 
Pushshift data has already been used to understand the phenomenon of hate speech and political extremism~\cite{extremism_deep_learning,chandrasekharan2017you,fair2019shouting,farrell2019exploring,grover2019detecting} and trolling and negative behaviors in fringe online spaces~\cite{almerekhi2019detecting,zannettou2018understanding,zannettou2018OriginsMemesMeans}.

\begin{table*}[t]
    \centering\small
    \begin{tabular}{l|cccc|c}
        \toprule[.15em]

        \rotatebox{45}{\textbf{Feature}} & 
        \rotatebox{45}{\textbf{Media Cloud}} & 
        \rotatebox{45}{\textbf{GDELT}} & 
        \rotatebox{45}{\textbf{StatsExchange}} & 
        \rotatebox{45}{\textbf{Wikimedia}} &
        \rotatebox{45}{\textbf{Pushshift (now)}}
        \\ \midrule
                
        \textit{Public API} & \halfcircle & \fullcircle & \fullcircle & \fullcircle & \fullcircle \\
        \textit{Data archive/dump} & \emptycircle & \fullcircle & \fullcircle & \fullcircle & \fullcircle\\
        \textit{Regularly updated} & \fullcircle & \fullcircle & \fullcircle & \fullcircle & \fullcircle\\
        \textit{Interactive computing} & \emptycircle & \halfcircle & \halfcircle & \fullcircle & \emptycircle\\
        \textit{Tutorials \& demos} & \emptycircle & \halfcircle & \fullcircle & \halfcircle & \halfcircle \\
        \textit{Online community} & \emptycircle & \emptycircle & \halfcircle & \halfcircle & \fullcircle\\
        \textit{Outreach} & \emptycircle & \halfcircle & \emptycircle & \halfcircle & \emptycircle \\
        \bottomrule
        
    \end{tabular}
    \caption{Features of big social data analysis cyber-infrastructures. \fullcircle~fully, \halfcircle~partially, and \emptycircle~not supported.}
    \label{tab:other_cyberinfrastructures}
\end{table*}

\paragraph{Online disinformation.} The online disinformation research community has focused its attention on how social media facilitates the spread of deliberately inaccurate information~\cite{computational_propaganda,starbird_icwsm}. 
The use of social media platforms to spread this ``fake news'' and biased political propaganda was particularly concerning given the events surrounding Russian interference in the 2016 US presidential election. 
Researchers studying disinformation acknowledge that mainstream platforms, particularly Facebook, are still the main place where disinformation campaigns take place~\cite{computational_propaganda2} and that a lack of data access is significantly limiting their efforts~\cite{davey_alba}.
While mainstream sites are the largest amplifiers of disinformation content, the content itself is often created on fringe sites that serve as proving grounds~\cite{poynter_gab,storyful_iceberg,marwick_and_lewis}. 
As with extremism and terrorism research, data access and data sharing in the disinformation research community is an ongoing struggle. 
Pushshift data has already been used in a number of papers on disinformation and social media trustworthiness~\cite{crothers2019towards,horne2017impact,zannettou2019let,zhou2019elites}.

\paragraph{Web science.} Datasets like Pushshift are critically important for researchers who answer questions at the intersection of Internet and society. 
How does technology spread? 
What is the impact of each interface or design choice on the efficacy of social media platforms? 
How should we measure the success or failure of an online community? 
Pushshift data has already been used in studies of user engagement on social media~\cite{aldous2019view}, social media moderation schemes~\cite{shen2019discourse,srinivasan2019content}, measuring success and growth of online communities~\cite{cunha2019all,tan2018tracing}, conflict in online groups~\cite{datta2019extracting,datta2017identifying,kumar2018community}, the spread of technological innovations~\cite{glenski2019characterizing}, modeling collaboration~\cite{kasper2017modeling,medvedev2018modelling}, and measuring engagement and collective attention~\cite{an2019political,lorenz2019accelerating}.

\paragraph{Big data science.} As one of a few easily-accessible, very large collections of social media data, Pushshift enables data-intensive research in foundational areas like network science~\cite{fire2019rise,sarantopoulos2018timerank,tsugawaimpact}, and new algorithms for cloud computing~\cite{kunft2018scootr} and very large databases~\cite{kunft2017blockjoin,kunft2019intermediate,ozcan2017bayesian}.

\paragraph{Health informatics.} Because of the relative anonymity allowed by certain social media platforms, large social media datasets are useful for researchers studying topics in health informatics including sensitive medical issues, atypical behaviors, and interpersonal conflict. 
Pushshift data has been used by researchers studying eating disorders and weight loss~\cite{enes2018reddit}, addiction and substance abuse~\cite{balsamo2019firsthand,barker2019topic,bowen2019increases,brett2019content,lu2019investigate,zhan2019underage}, sexually transmitted infections~\cite{lama2019characterizing}, difficult child-rearing problems~\cite{ammari2019self}, and various mental health challenges~\cite{chakravorti2018detecting,delahunty2018first,fraga2018online,grant2018automatic,grant2017discovery,pirina2018identifying,rezaii2019machine}.

\paragraph{Robust intelligence.}
Intelligent systems that can augment and enhance human understanding often require large amounts of human-generated text data generated in a social context. 
Social media data collected by Pushshift has been used already by researchers in computational linguistics and natural language processing~\cite{fulda2019semantically,gamallo2019contextualized,hidey2019fixed,jiang2018learning,wang2019can,zheng2019enhancing,zhuang2018quantifying}, recommender systems~\cite{buhagiar2018using,eberhard2019evaluating,halder2019predicting,hessel2017cats}, intelligent conversational systems~\cite{ahmadvand2019concet,golovanov2020lost,jonell2019crowdsourcing}, automatic summarization~\cite{volske2017tl}, entity recognition~\cite{derczynski2017results}, and other fields associated with the development of systems that can sense, reason, learn, and predict.

\section{Related Work}\label{sec:background}

\subsection{Existing Data Collection Services}

Promising alternatives to the aforementioned model of ``storage buckets of open data hosted by cloud providers'' exist that are better-tailored towards the needs of researchers.

Pushshift is not the first large-scale real-time social media data collection service aimed towards researchers.
Table~\ref{tab:other_cyberinfrastructures} summarizes the social and organizational features of other similar services.
While not an exhaustive list, the following have heavily influenced the research community as well as motivated Pushshift's own goals and design.

\begin{description}[leftmargin=1em]
    \item[Media Cloud] is an ``open source platform for studying media ecosystems'' that tracks hundreds of millions of news stories and makes aggregated count and topical data available via a free and semi-public API~\cite{chuang2014large}. The Media Cloud platform has been used to study digital health communication, agenda-setting, and online social movements. Researchers can use the API to get counts of stories, topics, words, and tags in response to queries by keyword, media source, and time window using a Solr search platform~\cite{mediacloud_api_2019}.
    \item[GDELT] is a free open platform monitoring global news media tracking events, related topics, and imagery. The platform offers a database and knowledge graph accessible both through dumps and an ``analysis service'' for filtering and visualizing subsets of the complete dataset~\cite{leetaru2013gdelt}. %
    \item[Stats Exchange] is a platform of social question answering communities, including Stack Overflow. While data dumps of the platform are hosted by the Internet Archive~\cite{stackexchange_dump_2019}, Stack Exchange offers both an API of activity as well as a ``Data Explorer'' allowing users to write SQL queries via a web interface against a regularly-updated database~\cite{stackexchange_explorer_2019}. %
    \item[Wikimedia] is the parent organization of projects like Wikipedia. It hosts data dumps of revision histories, content, and pageviews; makes data available through robust APIs; and offers a variety of interactive services. Wikimedia's deployment of Jupyter Notebooks can access replication databases of revisions and content. This enables researchers focus on analyzing data rather than system and database administration.
\end{description}

\noindent 

\descr{Other dataset papers.}
Considering the challenges in the post-API age, the collection, curation, and dissemination of datasets is crucial for the advancement of science.
To that end, it is worth exploring other works whose primary contribution has been the dataset they provide.
For example,~\cite{fair2019shouting} released a dataset that includes 37M posts and 24M comments covering August 2016 through December 2018 from Gab, a Twitter-like social media platform that after being de-platformed by major service providers ported their codebase to use the federated social network protocol from the Mastodon project.
As it turns out,~\cite{zignani2019mastodon} released a dataset focused around Mastodon itself.
Their dataset contains 5M posts, along with a crowdsourced (by Mastodon users) label that indicates whether or not the post contains inappropriate content.
Research into other types of computer-mediated communication platforms have also been enabled by dataset contributions.
~\cite{garimella2018whatapp} released a dataset from 178 WhatsApp groups that includes 454K messages from 45K different users.

\section{Discussion \& Conclusion}\label{sec:conclusion}

In this paper, we presented the Pushshift Reddit Dataset, which includes hundreds of millions of submissions and billions of comments from 2005 until the present.
In addition to offering Pushshift's data as monthly dumps, we also make this dataset available via a searchable API, as well as additional tools and community resources.
This paper also serves as a more formal and archival description of what Pushshift's Reddit dataset provides.
Having already been used in over 100 papers from numerous disciplines over the past four years, the Pushshift Reddit dataset will continue to be a valuable resource for the research community in the future.

\small
\bibliographystyle{abbrv}

\begin{thebibliography}{100}

\bibitem{ahmadvand2019concet}
A.~Ahmadvand, H.~Sahijwani, J.~I. Choi, and E.~Agichtein.
\newblock Concet: Entity-aware topic classification for open-domain
  conversational agents.
\newblock In {\em Proceedings of the 28th ACM International Conference on
  Information and Knowledge Management}, pages 1371--1380. ACM, 2019.

\bibitem{davey_alba}
D.~Alba.
\newblock Ahead of 2020, facebook falls short on plan to share data on
  disinformation.
\newblock
  \url{https://www.nytimes.com/2019/09/29/technology/facebook-disinformation.html},
  sep 2019.

\bibitem{aldous2019view}
K.~K. Aldous, J.~An, and B.~J. Jansen.
\newblock View, like, comment, post: Analyzing user engagement by topic at 4
  levels across 5 social media platforms for 53 news organizations.
\newblock In {\em Proceedings of the International AAAI Conference on Web and
  Social Media}, volume~13, pages 47--57, 2019.

\bibitem{almerekhi2019detecting}
H.~Almerekhi, H.~Kwak, B.~J. Jansen, and J.~Salminen.
\newblock Detecting toxicity triggers in online discussions.
\newblock In {\em Proceedings of the 30th ACM Conference on Hypertext and
  Social Media}, pages 291--292. ACM, 2019.

\bibitem{ammari2019self}
T.~Ammari, S.~Schoenebeck, and D.~Romero.
\newblock Self-declared throwaway accounts on reddit: How platform affordances
  and shared norms enable parenting disclosure and support.
\newblock {\em Proceedings of the ACM on Human-Computer Interaction},
  3(CSCW):135, 2019.

\bibitem{an2019political}
J.~An, H.~Kwak, O.~Posegga, and A.~Jungherr.
\newblock Political discussions in homogeneous and cross-cutting communication
  spaces.
\newblock In {\em Proceedings of the International AAAI Conference on Web and
  Social Media}, volume~13, pages 68--79, 2019.

\bibitem{stackexchange_dump_2019}
I.~Archive.
\newblock Stack exchange data dump.
\newblock \url{https://archive.org/details/stackexchange}, 2019.

\bibitem{balsamo2019firsthand}
D.~Balsamo, P.~Bajardi, and A.~Panisson.
\newblock Firsthand opiates abuse on social media: Monitoring geospatial
  patterns of interest through a digital cohort.
\newblock In {\em The World Wide Web Conference}, pages 2572--2579. ACM, 2019.

\bibitem{barker2019topic}
J.~O. Barker and J.~A. Rohde.
\newblock Topic clustering of e-cigarette submissions among reddit communities:
  A network perspective.
\newblock {\em Health Education \& Behavior}, 46(2\_suppl):59--68, 2019.

\bibitem{bastos}
M.~Bastos and S.~Walker.
\newblock Facebook's data lockdown is a disaster for academic researchers.
\newblock
  \url{https://theconversation.com/facebooks-data-lockdown-is-a-disaster-for-academic-researchers-94533},
  Apr. 2018.

\bibitem{baumgartner_incompleteness_2018}
J.~Baumgartner.
\newblock My response to the paper highlighting issues with data incompleteness
  concerning my reddit corpus.
\newblock \url{https://www.reddit.com/r/datasets/comments/884vkh/}, 2018.

\bibitem{Borgman_conundrumsharingresearch_2012}
C.~L. Borgman.
\newblock The conundrum of sharing research data.
\newblock {\em Journal of the American Society for Information Science and
  Technology}, 63(6):1059--1078, 2012.

\bibitem{botta}
A.~Botta, N.~Digiacomo, and K.~Mole.
\newblock Monetizing data: A new source of value in payments.
\newblock
  \url{https://www.mckinsey.com/industries/financial-services/our-insights/monetizing-
  data-a-new-source-of-value-in-payments}, 2017.

\bibitem{bowen2019increases}
D.~A. Bowen, J.~O’Donnell, and S.~A. Sumner.
\newblock Increases in online posts about synthetic opioids preceding increases
  in synthetic opioid death rates: a retrospective observational study.
\newblock {\em Journal of general internal medicine}, 34(12):2702--2704, 2019.

\bibitem{BoydUntanglingresearchpractice2016}
d.~boyd.
\newblock Untangling research and practice: {{What Facebook}}'s "emotional
  contagion" study teaches us.
\newblock {\em Research Ethics}, 12(1):4--13, 2016.

\bibitem{boydCriticalQuestionsBig2012}
d.~{boyd} and K.~Crawford.
\newblock Critical {{Questions}} for {{Big Data}}.
\newblock {\em Information, Communication \& Society}, 15(5):662--679, 2012.

\bibitem{boyle2017second}
J.~Boyle.
\newblock The second enclosure movement and the construction of the public
  domain.
\newblock In {\em Copyright Law}, pages 63--104. Routledge, 2017.

\bibitem{computational_propaganda2}
S.~Bradshaw and P.~N. Howard.
\newblock The global disinformation order: 2019 global inventory of organised
  social media manipulation.
\newblock
  \url{https://comprop.oii.ox.ac.uk/wp-content/uploads/sites/93/2019/09/CyberTroop-Report19.pdf},
  2019.

\bibitem{brett2019content}
E.~I. Brett, E.~M. Stevens, T.~L. Wagener, E.~L. Leavens, T.~L. Morgan, W.~D.
  Cotton, and E.~T. H{\'e}bert.
\newblock A content analysis of juul discussions on social media: Using reddit
  to understand patterns and perceptions of juul use.
\newblock {\em Drug and alcohol dependence}, 194:358--362, 2019.

\bibitem{Bruns_APIcalypsesocialmedia_2019}
A.~Bruns.
\newblock After the `{{APIcalypse}}': Social media platforms and their fight
  against critical scholarly research.
\newblock {\em Information, Communication \& Society}, 22(11):1544--1566, 2019.

\bibitem{buhagiar2018using}
N.~Buhagiar, B.~Zahir, and A.~Abhari.
\newblock Using deep learning to recommend discussion threads to users in an
  online forum.
\newblock In {\em 2018 International Joint Conference on Neural Networks
  (IJCNN)}, pages 1--8. IEEE, 2018.

\bibitem{burris_and_strahm}
V.~Burris, E.~Smith, and A.~Strahm.
\newblock White supremacist networks on the internet.
\newblock {\em Sociological Focus}, 33(2):215--235, 2000.

\bibitem{chakravorti2018detecting}
D.~Chakravorti, K.~Law, J.~Gemmell, and D.~Raicu.
\newblock Detecting and characterizing trends in online mental health
  discussions.
\newblock In {\em 2018 IEEE International Conference on Data Mining Workshops
  (ICDMW)}, pages 697--706. IEEE, 2018.

\bibitem{chandrasekharan_crossmod_2019}
E.~Chandrasekharan, C.~Gandhi, M.~W. Mustelier, and E.~Gilbert.
\newblock Crossmod: A cross-community learning-based system to assist reddit
  moderators.
\newblock {\em Proc. ACM Hum.-Comput. Interact.}, 3, Nov. 2019.

\bibitem{chandrasekharan2017you}
E.~Chandrasekharan, U.~Pavalanathan, A.~Srinivasan, A.~Glynn, J.~Eisenstein,
  and E.~Gilbert.
\newblock You can't stay here: The efficacy of reddit's 2015 ban examined
  through hate speech.
\newblock {\em Proceedings of the ACM on Human-Computer Interaction},
  1(CSCW):31, 2017.

\bibitem{chandrasekharan_ban_2017}
E.~Chandrasekharan, U.~Pavalanathan, A.~Srinivasan, A.~Glynn, J.~Eisenstein,
  and E.~Gilbert.
\newblock You can’t stay here: The efficacy of reddit’s 2015 ban examined
  through hate speech.
\newblock {\em Proc. ACM Hum.-Comput. Interact.}, 1, Dec. 2017.

\bibitem{chandrasekharan_hidden_2018}
E.~Chandrasekharan, M.~Samory, S.~Jhaver, H.~Charvat, A.~Bruckman, C.~Lampe,
  J.~Eisenstein, and E.~Gilbert.
\newblock The internet's hidden rules: An empirical study of reddit norm
  violations at micro, meso, and macro scales.
\newblock {\em Proc. ACM Hum.-Comput. Interact.}, 2, Nov. 2018.

\bibitem{chandrasekharan_bag_2017}
E.~Chandrasekharan, M.~Samory, A.~Srinivasan, and E.~Gilbert.
\newblock The bag of communities: Identifying abusive behavior online with
  preexisting internet data.
\newblock In {\em Proceedings of the 2017 CHI Conference on Human Factors in
  Computing Systems}, CHI ’17, page 3175–3187. {ACM}, 2017.

\bibitem{chuang2014large}
J.~Chuang, S.~Fish, D.~Larochelle, W.~P. Li, and R.~Weiss.
\newblock Large-scale topical analysis of multiple online news sources with
  media cloud.
\newblock In {\em NewsKDD: Data Science for News Publishing}, 2014.

\bibitem{mediacloud_api_2019}
M.~Cloud.
\newblock {API} specifications.
\newblock \url{https://github.com/berkmancenter/mediacloud/}, 2019.

\bibitem{ICU_2019}
I.~P.~M. Committee.
\newblock International components for unicode.
\newblock \url{http://site.icu-project.org/home}, 2019.

\bibitem{crothers2019towards}
E.~Crothers, N.~Japkowicz, and H.~L. Viktor.
\newblock Towards ethical content-based detection of online influence
  campaigns.
\newblock In {\em 2019 IEEE 29th International Workshop on Machine Learning for
  Signal Processing (MLSP)}, pages 1--6. IEEE, 2019.

\bibitem{cunha2019all}
T.~Cunha, D.~Jurgens, C.~Tan, and D.~Romero.
\newblock Are all successful communities alike? characterizing and predicting
  the success of online communities.
\newblock In {\em The World Wide Web Conference}, pages 318--328. ACM, 2019.

\bibitem{datta2019extracting}
S.~Datta and E.~Adar.
\newblock Extracting inter-community conflicts in reddit.
\newblock In {\em Proceedings of the International AAAI Conference on Web and
  Social Media}, volume~13, pages 146--157, 2019.

\bibitem{datta2017identifying}
S.~Datta, C.~Phelan, and E.~Adar.
\newblock Identifying misaligned inter-group links and communities.
\newblock {\em Proceedings of the ACM on Human-Computer Interaction},
  1(CSCW):37, 2017.

\bibitem{delahunty2018first}
F.~Delahunty, I.~D. Wood, and M.~Arcan.
\newblock First insights on a passive major depressive disorder prediction
  system with incorporated conversational chatbot.
\newblock In {\em AICS}, pages 327--338, 2018.

\bibitem{derczynski2017results}
L.~Derczynski, E.~Nichols, M.~van Erp, and N.~Limsopatham.
\newblock Results of the wnut2017 shared task on novel and emerging entity
  recognition.
\newblock In {\em Proceedings of the 3rd Workshop on Noisy User-generated
  Text}, pages 140--147, 2017.

\bibitem{eberhard2019evaluating}
L.~Eberhard, S.~Walk, L.~Posch, and D.~Helic.
\newblock Evaluating narrative-driven movie recommendations on reddit.
\newblock In {\em IUI}, pages 1--11, 2019.

\bibitem{Ekbia_Bigdatabigger_2015}
H.~Ekbia, M.~Mattioli, I.~Kouper, G.~Arave, A.~Ghazinejad, T.~Bowman, V.~R.
  Suri, A.~Tsou, S.~Weingart, and C.~R. Sugimoto.
\newblock Big data, bigger dilemmas: {{A}} critical review.
\newblock {\em Journal of the Association for Information Science and
  Technology}, 66(8):1523--1545, 2015.

\bibitem{ICU_ES_2019}
Elasticsearch.
\newblock Icu analysis plugin.
\newblock
  \url{https://www.elastic.co/guide/en/elasticsearch/plugins/current/analysis-icu.html},
  2019.

\bibitem{enes2018reddit}
K.~B. Enes, P.~P.~V. Brum, T.~O. Cunha, F.~Murai, A.~P.~C. da~Silva, and G.~L.
  Pappa.
\newblock Reddit weight loss communities: do they have what it takes for
  effective health interventions?
\newblock In {\em 2018 IEEE/WIC/ACM International Conference on Web
  Intelligence (WI)}, pages 508--513. IEEE, 2018.

\bibitem{stackexchange_explorer_2019}
S.~Exchange.
\newblock Data explorer help.
\newblock \url{https://data.stackexchange.com/help}, 2019.

\bibitem{fair2019shouting}
G.~Fair and R.~Wesslen.
\newblock Shouting into the void: A database of the alternative social media
  platform gab.
\newblock In {\em Proceedings of the International AAAI Conference on Web and
  Social Media}, volume~13, pages 608--610, 2019.

\bibitem{farrell2019exploring}
T.~Farrell, M.~Fernandez, J.~Novotny, and H.~Alani.
\newblock Exploring misogyny across the manosphere in reddit.
\newblock In {\em Proceedings of the 10th ACM Conference on Web Science},
  WebSci'19, pages 87--96, 2019.

\bibitem{fiesler2018reddit}
C.~Fiesler, J.~McCann, K.~Frye, J.~R. Brubaker, et~al.
\newblock Reddit rules! characterizing an ecosystem of governance.
\newblock In {\em Twelfth International AAAI Conference on Web and Social
  Media}. {AAAI}, 2018.

\bibitem{fire2019rise}
M.~Fire and C.~Guestrin.
\newblock The rise and fall of network stars: Analyzing 2.5 million graphs to
  reveal how high-degree vertices emerge over time.
\newblock {\em Information Processing \& Management}, 2019.

\bibitem{wikimedia_dump_2019}
W.~Foundation.
\newblock Wikimedia downloads.
\newblock \url{https://dumps.wikimedia.org/}, 2019.

\bibitem{fraga2018online}
B.~S. Fraga, A.~P.~C. da~Silva, and F.~Murai.
\newblock Online social networks in health care: A study of mental disorders on
  reddit.
\newblock In {\em 2018 IEEE/WIC/ACM International Conference on Web
  Intelligence (WI)}, pages 568--573. IEEE, 2018.

\bibitem{Freelon_InterpretationDigitalTrace_2014}
D.~Freelon.
\newblock On the {{Interpretation}} of {{Digital Trace Data}} in
  {{Communication}} and {{Social Computing Research}}.
\newblock {\em Journal of Broadcasting \& Electronic Media}, 58(1):59--75,
  2014.

\bibitem{Freelon_ComputationalResearchPostAPI_2018}
D.~Freelon.
\newblock Computational {{Research}} in the {{Post}}-{{API Age}}.
\newblock {\em Political Communication}, 35(4):665--668, 2018.

\bibitem{fulda2019semantically}
N.~E. Fulda.
\newblock Semantically aligned sentence-level embeddings for agent autonomy and
  natural language understanding.
\newblock 2019.

\bibitem{poynter_gab}
D.~Funke.
\newblock Misinformers are moving to smaller platforms. so how should
  fact-checkers monitor them?
\newblock
  \url{https://www.poynter.org/fact-checking/2018/misinformers-are-moving-to-smaller-platforms-so-how-should-fact-
  checkers-monitor-them/}, Dec. 2018.

\bibitem{gaffney_caveat_2018}
D.~Gaffney and J.~N. Matias.
\newblock Caveat emptor, computational social science: {{Large}}-scale missing
  data in a widely-published {{Reddit}} corpus.
\newblock {\em PLOS ONE}, 13(7):e0200162, July 2018.

\bibitem{gamallo2019contextualized}
P.~Gamallo, S.~Sotelo, J.~R. Pichel, and M.~Artetxe.
\newblock Contextualized translations of phrasal verbs with distributional
  compositional semantics and monolingual corpora.
\newblock {\em Computational Linguistics}, pages 1--27, 2019.

\bibitem{garimella2018whatapp}
K.~Garimella and G.~Tyson.
\newblock {Whatapp Doc? A First Look at Whatsapp Public Group Data}.
\newblock In {\em Twelfth International AAAI Conference on Web and Social
  Media}, 2018.

\bibitem{gibney_privacy_2019}
E.~Gibney.
\newblock Privacy hurdles thwart {{Facebook}} democracy research.
\newblock {\em Nature}, 574(7777):158--159, Oct. 2019.

\bibitem{glenski2019characterizing}
M.~Glenski, E.~Saldanha, and S.~Volkova.
\newblock Characterizing speed and scale of cryptocurrency discussion spread on
  reddit.
\newblock In {\em The World Wide Web Conference}, pages 560--570. ACM, 2019.

\bibitem{GolderDigitalFootprintsOpportunities2014}
S.~A. Golder and M.~W. Macy.
\newblock Digital {{Footprints}}: {{Opportunities}} and {{Challenges}} for
  {{Online Social Research}}.
\newblock {\em Annual Review of Sociology}, 40(1):129--152, 2014.

\bibitem{golovanov2020lost}
S.~Golovanov, A.~Tselousov, R.~Kurbanov, and S.~I. Nikolenko.
\newblock Lost in conversation: A conversational agent based on the transformer
  and transfer learning.
\newblock In {\em The NeurIPS'18 Competition}, pages 295--315. Springer, 2020.

\bibitem{storyful_iceberg}
D.~Gonimah.
\newblock Storyful's guide to the social media landscape: Beyond the iceberg
  metaphor.
\newblock
  \url{https://storyful.com/thought-leadership/storyfuls-guide-to-the-social-media-landscape-beyond-the-iceberg-
  metaphor/}, oct 2018.

\bibitem{Gousi13}
G.~Gousios.
\newblock The ghtorrent dataset and tool suite.
\newblock In {\em Proceedings of the 10th Working Conference on Mining Software
  Repositories}, MSR '13. IEEE Press, 2013.

\bibitem{grant2017discovery}
R.~Grant, D.~Kucher, A.~M. Le{\'o}n, J.~Gemmell, and D.~Raicu.
\newblock Discovery of informal topics from post traumatic stress disorder
  forums.
\newblock In {\em 2017 IEEE International Conference on Data Mining Workshops
  (ICDMW)}, pages 452--461. IEEE, 2017.

\bibitem{grant2018automatic}
R.~N. Grant, D.~Kucher, A.~M. Le{\'o}n, J.~F. Gemmell, D.~S. Raicu, and S.~J.
  Fodeh.
\newblock Automatic extraction of informal topics from online suicidal
  ideation.
\newblock {\em BMC bioinformatics}, 19(8):211, 2018.

\bibitem{grover2019detecting}
T.~Grover and G.~Mark.
\newblock Detecting potential warning behaviors of ideological radicalization
  in an alt-right subreddit.
\newblock In {\em Proceedings of the International AAAI Conference on Web and
  Social Media}, volume~13, pages 193--204, 2019.

\bibitem{Halavais_Overcomingtermsservice_2019}
A.~Halavais.
\newblock Overcoming terms of service: A proposal for ethical distributed
  research.
\newblock {\em Information, Communication \& Society}, 22(11):1567--1581, 2019.

\bibitem{halder2019predicting}
K.~Halder, M.-Y. Kan, and K.~Sugiyama.
\newblock Predicting helpful posts in open-ended discussion forums: A neural
  architecture.
\newblock In {\em Proceedings of the 2019 Conference of the North American
  Chapter of the Association for Computational Linguistics: Human Language
  Technologies, Volume 1 (Long and Short Papers)}, pages 3148--3157, 2019.

\bibitem{Hampton_StudyingDigitalDirections_2017}
K.~N. Hampton.
\newblock Studying the {{Digital}}: {{Directions}} and {{Challenges}} for
  {{Digital Methods}}.
\newblock {\em Annual Review of Sociology}, 43(1):167--188, 2017.

\bibitem{hess2003ideas}
C.~Hess and E.~Ostrom.
\newblock Ideas, artifacts, and facilities: information as a common-pool
  resource.
\newblock {\em Law and contemporary problems}, 66(1/2):111--145, 2003.

\bibitem{hessel2017cats}
J.~Hessel, L.~Lee, and D.~Mimno.
\newblock Cats and captions vs. creators and the clock: Comparing multimodal
  content to context in predicting relative popularity.
\newblock In {\em Proceedings of the 26th International Conference on World
  Wide Web}, pages 927--936. International World Wide Web Conferences Steering
  Committee, 2017.

\bibitem{hidey2019fixed}
C.~Hidey and K.~McKeown.
\newblock Fixed that for you: Generating contrastive claims with semantic
  edits.
\newblock In {\em Proceedings of the 2019 Conference of the North American
  Chapter of the Association for Computational Linguistics: Human Language
  Technologies, Volume 1 (Long and Short Papers)}, pages 1756--1767, 2019.

\bibitem{horne2017impact}
B.~D. Horne and S.~Adali.
\newblock The impact of crowds on news engagement: A reddit case study.
\newblock In {\em Eleventh International AAAI Conference on Web and Social
  Media}, 2017.

\bibitem{howison_validity_2011}
J.~Howison, A.~Wiggins, and K.~Crowston.
\newblock Validity {{Issues}} in the {{Use}} of {{Social Network Analysis}}
  with {{Digital Trace Data}}.
\newblock {\em Journal of the Association for Information Systems; Atlanta},
  12(12):767--797, Dec. 2011.

\bibitem{hunter2003cyberspace}
D.~Hunter.
\newblock Cyberspace as place and the tragedy of the digital anticommons.
\newblock {\em California Law Review}, 91:439, 2003.

\bibitem{ingram_silicon_2019}
M.~Ingram.
\newblock Silicon {Valley}'s {Stonewalling}.
\newblock {\em Columbia Journalism Review}, 2019.

\bibitem{Japec_BigDataSurvey_2015}
L.~Japec, F.~Kreuter, M.~Berg, P.~Biemer, P.~Decker, C.~Lampe, J.~Lane,
  C.~O'Neil, and A.~Usher.
\newblock Big {{Data}} in {{Survey Research}}: {{AAPOR Task Force Report}}.
\newblock {\em Public Opinion Quarterly}, 79(4):839--880, 2015.

\bibitem{jhaver_reactions_2019}
S.~Jhaver, D.~S. Appling, E.~Gilbert, and A.~Bruckman.
\newblock ``did you suspect the post would be removed?'': Understanding user
  reactions to content removals on reddit.
\newblock {\em Proc. ACM Hum.-Comput. Interact.}, 3, Nov. 2019.

\bibitem{jhaver_transparency_2019}
S.~Jhaver, A.~Bruckman, and E.~Gilbert.
\newblock Does transparency in moderation really matter? user behavior after
  content removal explanations on reddit.
\newblock {\em Proc. ACM Hum.-Comput. Interact.}, 3, Nov. 2019.

\bibitem{jiang2018learning}
J.-Y. Jiang, F.~Chen, Y.-Y. Chen, and W.~Wang.
\newblock Learning to disentangle interleaved conversational threads with a
  siamese hierarchical network and similarity ranking.
\newblock In {\em Proceedings of the 2018 Conference of the North American
  Chapter of the Association for Computational Linguistics: Human Language
  Technologies, Volume 1 (Long Papers)}, pages 1812--1822, 2018.

\bibitem{extremism_deep_learning}
A.~Johnston and A.~Marku.
\newblock Identifying extremism in text using deep learning.
\newblock {\em Development and Analysis of Deep Learning Architectures}, pages
  267--289, 2020.

\bibitem{jonell2019crowdsourcing}
P.~Jonell, P.~Fallgren, F.~I. Do{\u{g}}an, J.~Lopes, U.~Wennberg, and
  G.~Skantze.
\newblock Crowdsourcing a self-evolving dialog graph.
\newblock In {\em Proceedings of the 1st International Conference on
  Conversational User Interfaces}, page~14. ACM, 2019.

\bibitem{kasper2017modeling}
P.~Kasper, P.~Koncar, S.~Walk, T.~Santos, M.~W{\"o}lbitsch, M.~Strohmaier, and
  D.~Helic.
\newblock Modeling user dynamics in collaboration websites.
\newblock In {\em Dynamics on and of Complex Networks}, pages 113--133.
  Springer, 2017.

\bibitem{Keegan_DiscoveringSocial_2018}
B.~Keegan.
\newblock Discovering the {{Social}}.
\newblock \url{http://www.brianckeegan.com/2018/03/discovering-the-social/},
  2018.

\bibitem{kumar2018community}
S.~Kumar, W.~L. Hamilton, J.~Leskovec, and D.~Jurafsky.
\newblock Community interaction and conflict on the web.
\newblock In {\em Proceedings of the 2018 World Wide Web Conference}, pages
  933--943. International World Wide Web Conferences Steering Committee, 2018.

\bibitem{kunft2019intermediate}
A.~Kunft, A.~Katsifodimos, S.~Schelter, S.~Bre{\ss}, T.~Rabl, and V.~Markl.
\newblock An intermediate representation for optimizing machine learning
  pipelines.
\newblock {\em Proceedings of the VLDB Endowment}, 12(11):1553--1567, 2019.

\bibitem{kunft2017blockjoin}
A.~Kunft, A.~Katsifodimos, S.~Schelter, T.~Rabl, and V.~Markl.
\newblock Blockjoin: efficient matrix partitioning through joins.
\newblock {\em Proceedings of the VLDB Endowment}, 10(13):2061--2072, 2017.

\bibitem{kunft2018scootr}
A.~Kunft, L.~Stadler, D.~Bonetta, C.~Basca, J.~Meiners, S.~Bre{\ss}, T.~Rabl,
  J.~Fumero, and V.~Markl.
\newblock Scootr: Scaling r dataframes on dataflow systems.
\newblock In {\em Proceedings of the ACM Symposium on Cloud Computing}, pages
  288--300. ACM, 2018.

\bibitem{lama2019characterizing}
Y.~Lama, D.~Hu, A.~Jamison, S.~C. Quinn, and D.~A. Broniatowski.
\newblock Characterizing trends in human papillomavirus vaccine discourse on
  reddit (2007-2015): An observational study.
\newblock {\em JMIR public health and surveillance}, 5(1):e12480, 2019.

\bibitem{LazerComputationalSocialScience2009}
D.~Lazer, A.~Pentland, L.~Adamic, S.~Aral, A.-L. Barab{\'a}si, D.~Brewer,
  N.~Christakis, N.~Contractor, J.~Fowler, M.~Gutmann, T.~Jebara, G.~King,
  M.~Macy, D.~Roy, and M.~V. Alstyne.
\newblock Computational {{Social Science}}.
\newblock {\em Science}, 323(5915):721--723, 2009.

\bibitem{Lazer_DataexMachina_2017}
D.~Lazer and J.~Radford.
\newblock Data ex {{Machina}}: {{Introduction}} to {{Big Data}}.
\newblock {\em Annual Review of Sociology}, 43(1):19--39, 2017.

\bibitem{leetaru2013gdelt}
K.~Leetaru and P.~A. Schrodt.
\newblock Gdelt: Global data on events, location, and tone, 1979--2012.
\newblock In {\em ISA annual convention}, number~4, pages 1--49, 2013.

\bibitem{lorenz2019accelerating}
P.~Lorenz-Spreen, B.~M. M{\o}nsted, P.~H{\"o}vel, and S.~Lehmann.
\newblock Accelerating dynamics of collective attention.
\newblock {\em Nature communications}, 10(1):1759, 2019.

\bibitem{lu2019investigate}
J.~Lu, S.~Sridhar, R.~Pandey, M.~A. Hasan, and G.~Mohler.
\newblock Investigate transitions into drug addiction through text mining of
  reddit data.
\newblock In {\em Proceedings of the 25th ACM SIGKDD International Conference
  on Knowledge Discovery \& Data Mining}, pages 2367--2375. ACM, 2019.

\bibitem{manovich2011trending}
L.~Manovich.
\newblock Trending: The promises and the challenges of big social data.
\newblock {\em Debates in the digital humanities}, 2:460--475, 2011.

\bibitem{marwick_and_lewis}
A.~Marwick and R.~Lewis.
\newblock Media manipulation and disinformation online: Case studies.
\newblock
  \url{https://datasociety.net/pubs/oh/DataAndSociety\_MediaManipulationAndDisinformationOnline.pdf},
  may 2017.

\bibitem{massanari_gamergate_2017}
A.~Massanari.
\newblock \#gamergate and the fappening: How reddit’s algorithm, governance,
  and culture support toxic technocultures.
\newblock {\em New Media \& Society}, 19(3):329--346, 2017.

\bibitem{matias_dark_2016}
J.~N. Matias.
\newblock Going dark: Social factors in collective action against platform
  operators in the reddit blackout.
\newblock In {\em Proceedings of the 2016 CHI Conference on Human Factors in
  Computing Systems}, CHI ’16, page 1138–1151. {ACM}, 2016.

\bibitem{matias_civic_2019}
J.~N. Matias.
\newblock The {{Civic Labor}} of {{Volunteer Moderators Online}}.
\newblock {\em Social Media + Society}, Apr. 2019.

\bibitem{medvedev2018modelling}
A.~N. Medvedev, J.-C. Delvenne, and R.~Lambiotte.
\newblock Modelling structure and predicting dynamics of discussion threads in
  online boards.
\newblock {\em Journal of Complex Networks}, 7(1):67--82, 2018.

\bibitem{Mervis_Privacyconcernscould_2019}
J.~Mervis.
\newblock Privacy concerns could derail {{Facebook}} data-sharing plan.
\newblock {\em Science}, 365(6460):1360--1361, 2019.

\bibitem{computational_propaganda}
V.~Narayanan, V.~Barash, J.~Kelly, B.~Kollanyi, L.-M. Neudert, and P.~N.
  Howard.
\newblock Polarization, partisanship and junk news consumption over the us.
\newblock
  \url{http://comprop.oii.ox.ac.uk/wp-content/uploads/sites/93/2018/02/Polarization-Partisanship-JunkNews.pdf},
  2018.

\bibitem{onlinehateprevention}
A.~Oboloer, W.~Allington, and P.~Scolyer-Gray.
\newblock Hate and violent extremism from an online subculture: The yom kippur
  terrorist attack in halle, germany.
\newblock
  \url{http://ohpi.org.au/halle/Hate\%20and\%20Violent\%20Extremism\%20from\%20an\%20Online\%20Sub-Culture.pdf},
  2019.

\bibitem{Olteanu_SocialDataBiases_2019}
A.~Olteanu, C.~Castillo, F.~Diaz, and E.~K\i{}c\i{}man.
\newblock Social {{Data}}: {{Biases}}, {{Methodological Pitfalls}}, and
  {{Ethical Boundaries}}.
\newblock {\em Frontiers in Big Data}, 2, 2019.

\bibitem{ozcan2017bayesian}
F.~Ozcan.
\newblock {\em Bayesian Nonparametric Models on Big Data}.
\newblock PhD thesis, UC Irvine, 2017.

\bibitem{Palen_CrisisinformaticsNew_2016}
L.~Palen and K.~M. Anderson.
\newblock Crisis informatics\textemdash{{New}} data for extraordinary times.
\newblock {\em Science}, 353(6296):224--225, 2016.

\bibitem{Patel_TestingLimitsFirst_2018}
K.~S. Patel.
\newblock Testing the {{Limits}} of the {{First Amendment}}: {{How Online Civil
  Rights Testing}} is {{Protected Speech Activity}}.
\newblock {\em Columbia Law Review}, 118(5):1473--1516, 2018.

\bibitem{pirina2018identifying}
I.~Pirina and {\c{C}}.~{\c{C}}{\"o}ltekin.
\newblock Identifying depression on reddit: The effect of training data.
\newblock In {\em Proceedings of the 2018 EMNLP Workshop SMM4H: The 3rd Social
  Media Mining for Health Applications Workshop \& Shared Task}, pages 9--12,
  2018.

\bibitem{Puschmann_endwildwest_2019}
C.~Puschmann.
\newblock An end to the wild west of social media research: A response to
  {{Axel Bruns}}.
\newblock {\em Information, Communication \& Society}, 22(11):1582--1589, 2019.

\bibitem{reddit_api_2019}
Reddit.
\newblock {API} documentation.
\newblock \url{https://www.reddit.com/dev/api/}, 2019.

\bibitem{rezaii2019machine}
N.~Rezaii, E.~Walker, and P.~Wolff.
\newblock A machine learning approach to predicting psychosis using semantic
  density and latent content analysis.
\newblock {\em NPJ schizophrenia}, 5, 2019.

\bibitem{sarantopoulos2018timerank}
I.~Sarantopoulos, D.~Papatheodorou, D.~Vogiatzis, G.~Tzortzis, and
  G.~Paliouras.
\newblock Timerank: A random walk approach for community discovery in dynamic
  networks.
\newblock In {\em International Conference on Complex Networks and their
  Applications}, pages 338--350. Springer, 2018.

\bibitem{seering_moderator_2019}
J.~Seering, T.~Wang, J.~Yoon, and G.~Kaufman.
\newblock Moderator engagement and community development in the age of
  algorithms.
\newblock {\em New Media \& Society}, 21(7):1417--1443, July 2019.

\bibitem{shen2019discourse}
Q.~Shen and C.~Rose.
\newblock The discourse of online content moderation: Investigating polarized
  user responses to changes in reddit’s quarantine policy.
\newblock In {\em Proceedings of the Third Workshop on Abusive Language
  Online}, pages 58--69, 2019.

\bibitem{squirrell_platform_2019}
T.~Squirrell.
\newblock Platform dialectics: {{The}} relationships between volunteer
  moderators and end users on reddit.
\newblock {\em New Media \& Society}, Mar. 2019.

\bibitem{srinivasan2019content}
K.~B. Srinivasan, C.~Danescu-Niculescu-Mizil, L.~Lee, and C.~Tan.
\newblock Content removal as a moderation strategy: Compliance and other
  outcomes in the changemyview community.
\newblock {\em Proceedings of the ACM on Human-Computer Interaction},
  3(CSCW):163, 2019.

\bibitem{starbird_icwsm}
K.~Starbird.
\newblock Examining the alternative media ecosystem through the production of
  alternative narratives of mass shooting events on twitter.
\newblock In {\em International AAAI Conference on Web and Social Media}.
  {AAAI}, 2017.

\bibitem{tan2018tracing}
C.~Tan.
\newblock Tracing community genealogy: how new communities emerge from the old.
\newblock In {\em Twelfth International AAAI Conference on Web and Social
  Media}, 2018.

\bibitem{tech_versus_terrorism}
T.~A. Terrorism.
\newblock Insights from the centre for analysis of the radical right’s
  inaugural conference in london.
\newblock
  \url{https://www.techagainstterrorism.org/2019/06/06/insights-from-the-centre-for-analysis-of-the-radical-rights-
  inaugural-conference-in-london/}, May 2019.

\bibitem{tsugawaimpact}
S.~Tsugawa and S.~Niida.
\newblock The impact of social network structure on the growth and survival of
  online communities.
\newblock {\em International Conference on Advances in Social Networks Analysis
  and Mining}, pages 1112--1119, 2019.

\bibitem{tufekci2014big}
Z.~Tufekci.
\newblock Big questions for social media big data: Representativeness, validity
  and other methodological pitfalls.
\newblock In {\em Eighth International AAAI Conference on Weblogs and Social
  Media}. {AAAI}, 2014.

\bibitem{volske2017tl}
M.~V{\"o}lske, M.~Potthast, S.~Syed, and B.~Stein.
\newblock Tl; dr: Mining reddit to learn automatic summarization.
\newblock In {\em Proceedings of the Workshop on New Frontiers in
  Summarization}, pages 59--63, 2017.

\bibitem{Walker_disinformationlandscapelockdown_2019}
S.~Walker, D.~Mercea, and M.~Bastos.
\newblock The disinformation landscape and the lockdown of social platforms.
\newblock {\em Information, Communication \& Society}, 22(11):1531--1543, 2019.

\bibitem{wang2019can}
A.~Wang, J.~Hula, P.~Xia, R.~Pappagari, R.~T. McCoy, R.~Patel, N.~Kim,
  I.~Tenney, Y.~Huang, K.~Yu, et~al.
\newblock Can you tell me how to get past sesame street? sentence-level
  pretraining beyond language modeling.
\newblock In {\em Proceedings of the 57th Annual Meeting of the Association for
  Computational Linguistics}, pages 4465--4476, 2019.

\bibitem{wang_social_2007}
F.-Y. Wang, K.~M. Carley, D.~Zeng, and W.~Mao.
\newblock Social {{Computing}}: {{From Social Informatics}} to {{Social
  Intelligence}}.
\newblock {\em IEEE Intelligent Systems}, 22(2):79--83, Mar. 2007.

\bibitem{Weller_manifestodatasharing_2016}
K.~Weller and K.~E. {Kinder-Kurlanda}.
\newblock A manifesto for data sharing in social media research.
\newblock In {\em Proceedings of the 8th {{ACM Conference}} on {{Web
  Science}}}, pages 166--172. {ACM}, 2016.

\bibitem{zannettou2018understanding}
S.~Zannettou, J.~Blackburn, E.~De~Cristofaro, M.~Sirivianos, and G.~Stringhini.
\newblock Understanding web archiving services and their (mis) use on social
  media.
\newblock In {\em Twelfth International AAAI Conference on Web and Social
  Media}, 2018.

\bibitem{zannettou2018OriginsMemesMeans}
S.~Zannettou, T.~Caulfield, J.~Blackburn, E.~De~Cristofaro, M.~Sirivianos,
  G.~Stringhini, and G.~{Suarez-Tangil}.
\newblock On the {{Origins}} of {{Memes}} by {{Means}} of {{Fringe Web
  Communities}}.
\newblock In {\em Proceedings of the {{Internet Measurement Conference}} 2018},
  {{IMC}} '18, pages 188--202, {Boston, MA, USA}, 2018. {ACM}.

\bibitem{zannettou2019let}
S.~Zannettou, T.~Caulfield, W.~Setzer, M.~Sirivianos, G.~Stringhini, and
  J.~Blackburn.
\newblock Who let the trolls out?: Towards understanding state-sponsored
  trolls.
\newblock In {\em Proceedings of the 10th ACM Conference on Web Science}, pages
  353--362. ACM, 2019.

\bibitem{zhan2019underage}
Y.~Zhan, Z.~Zhang, J.~M. Okamoto, D.~D. Zeng, and S.~J. Leischow.
\newblock Underage juul use patterns: Content analysis of reddit messages.
\newblock {\em Journal of medical Internet research}, 21(9):e13038, 2019.

\bibitem{zheng2019enhancing}
W.~Zheng and K.~Zhou.
\newblock Enhancing conversational dialogue models with grounded knowledge.
\newblock In {\em Proceedings of the 28th ACM International Conference on
  Information and Knowledge Management}, pages 709--718. ACM, 2019.

\bibitem{zhou2019elites}
Y.~Zhou, M.~Dredze, D.~A. Broniatowski, and W.~D. Adler.
\newblock Elites and foreign actors among the alt-right: The gab social media
  platform.
\newblock {\em First Monday}, 24(9), 2019.

\bibitem{zhuang2018quantifying}
Y.~Zhuang, J.~Xie, Y.~Zheng, and X.~Zhu.
\newblock Quantifying context overlap for training word embeddings.
\newblock In {\em Proceedings of the 2018 Conference on Empirical Methods in
  Natural Language Processing}, pages 587--593, 2018.

\bibitem{zignani2019mastodon}
M.~Zignani, C.~Quadri, A.~Galdeman, S.~Gaito, and G.~P. Rossi.
\newblock {Mastodon Content Warnings: Inappropriate Contents in a Microblogging
  Platform}.
\newblock In {\em Proceedings of the International AAAI Conference on Web and
  Social Media}, volume~13, pages 639--645, 2019.

\end{thebibliography}

\end{document}